\documentclass[useAMS,usenatbib]{mn2e}
\usepackage{graphicx}
\usepackage{journals}
\renewcommand{\epsilon}{\varepsilon}
\renewcommand{\phi}{\varphi}
\newcommand{\msun}{\mbox{${\rm M}_\odot$}}
\newcommand{\nbody}{\emph{N}--body~}

\def\cite#1{\citealt{#1}}
\newcommand {\apgt} {\ {\raise-.5ex\hbox{$\buildrel>\over\sim$}}\ }
\newcommand {\aplt} {\ {\raise-.5ex\hbox{$\buildrel<\over\sim$}}\ }

\begin{document}
%
%\title[The multiple star reaction network in small star clusters]{The
%multiple star reaction network in small star clusters}
\title[The multiple star reaction network in small star clusters]{The
formation of higher-order hierarchical systems in star clusters}
\author[J. van den Berk, S. F. Portegies Zwart, S. L. W. McMillan]{J. van den Berk$^{1}
$, S. F. Portegies Zwart$^{1,2}$\thanks{spz@science.uva.nl}, S. L. W. McMillan$^{3}$\\
$^{1}$Astronomical Institute 'Anton Pannekoek', University of Amsterdam, Amsterdam, the Netherlands\\
$^{2}$Computational Science Section, University of Amsterdam, Amsterdam, the Netherlands\\
$^{3}$Department of Physics, Drexel University, Philadelphia, PA 19104, USA}
\date{Accepted 2006 ???. Received 2006 ???; in original form 2006 ???}
\pagerange{\pageref{firstpage}--\pageref{lastpage}} \pubyear{2006}
\maketitle
\label{firstpage}
\begin{abstract}
We simulate open clusters containing up to 182 stars initially in the
form of singles, binaries and triples.  Due to the high interaction
rate a large number of stable quadruples, quintuples, sextuples, and
higher-order hierarchies form during the course of the simulations.
For our choice of initial conditions, the formation rate of quadruple
systems after about 2\,Myr is roughly constant with time at $\sim
0.008$ per cluster per Myr.  The formation rate of quintuple and
sextuple systems are about half and one quarter, respectively, of the
quadruple formation rate, and both rates are also approximately
constant with time.  We present reaction channels and relative
probabilities for the formation of persistent systems containing up to
six stars.  The reaction networks for the formation and destruction of
quintuple and sextuple systems can become quite complicated, although
the branching ratios remain largely unchanged during the course of the
cluster evolution.  The total number of quadruples is about a factor
of three smaller than observed in the solar neighbourhood.
\end{abstract}

\section{Introduction}\label{sec:intro}
A sizable fraction of (and possibly all) stars are formed in stellar
clusters (Lada \& Lada 2003, de Grijd et al.\, 2003, Fall et al.\,
2005).  \nocite{2003ARA&A..41...57L,GL2003b,FC2005} Investigations of
starburst regions in other galaxies reveal that star clusters are
predominantly of low mass.  For the small Magellanic cloud and the
galaxies M33 and M51, \cite{LG2005} find that the initial cluster
mass function follows a power law distribution with a slope of about
-2. The minimum mass in their sample ($M_{\rm cl} \simeq 10^4$\,\msun)
is mainly a result of observational bias.  The mean mass of star
clusters in the solar neighbourhood is $\langle M_{\rm cl} \rangle
\simeq 1000$\,\msun\,\citep{KP2005}, suggesting that the mass function
found by \citet{GA2003a} should be extended to even lower masses. The
vast majority ($96\pm2$\%) of stars of spectral type O do not belong
to a known association, and may have formed in small star clusters
\citep{DT2005} and subsequently been ejected by dynamical slingshots
\citep{GZ2004,AA2005}.  From this it has been argued that the majority
of stars are born in small clusters \citep{KR1995a}, many of which may
disperse within a few tens of Myr of their formation \citep{BG2005,
FC2005}.  In this case, the population of Galactic field stars
originates mainly from low-mass clusters and, as a consequence, the
majority of stellar multiples in the Galactic field were also born in
relatively small star clusters.

The population of binary stars in the solar neighbourhood has been
studied extensively for the field \citep{DM1991,C1993,CO1995} and for
some nearby star clusters \citep{KB2005}.  Higher-order
multiplicities are much more difficult to find than binaries.  Still,
the Multiple Star Catalogue (MSC) contains 728 systems comprising 3--7
stars.  The catalogue is claimed to be complete to a distance of
{10}\,pc from the Sun \citep{MSC1997, MSC1999}. The majority of the
listed systems are triples, and orbital parameters are provided where
they are available \citep{ST2002}.  The MSC contains 558 triples, 138
quadruples, 25 quintuples, and 7 sextuples.

We define the multiplicity fraction as the number of objects with a
given multiplicity divided by the total number of single stars and
multiples:
\[
	f_{\rm i} = n_i/N,
\]
Here $n_i$ is the number of objects with multiplicity $i$ and $N$ is
the sum of the number of single stars ($N_{\rm S}$), binaries ($N_{\rm
B}$), triples ($N_{\rm T}$), quadruples ($N_{\rm Qd}$), quintuples
($N_{\rm Qn}$) and sextuples ($N_{\rm Sx}$).  Assuming that 15--25\%
of all systems have multiplicity larger than two \citep{TO2004}, we
arrive at the following multiplicity fractions: $f_{\rm T} =
0.11$--0.19, $f_{\rm Qd} = 0.03$--0.05, $f_{\rm Qn} = 0.005$--0.009
and $f_{\rm Sx} = 0.001$--0.002.  (A summary of the multiplicities
found in the MSC and their configuration is presented in
Table\,\ref{tab:multipletypes}.)

If most stars are born in relatively small star clusters, then the
majority of the field population must originate in small clusters and
the proportions in which single stars, binaries, triples and higher
order multiples occur in the field should be reproducible by computer
simulations of such star clusters.  Once these clusters dissolve, the
numbers of higher-order multiples are no longer affected by cluster
dynamical evolution, although internal stellar evolution may still
affect their relative numbers.  For example, a triple can evolve into
a binary when two of its components merge in an unstable phase of
Roche-lobe overflow.  Thus, the relative multiplicity fraction is not
frozen in when the cluster dissolves.  Stellar evolution tends to
reduce the multiplicity, although these effects become less important
for older populations.

Several recent theoretical studies have discussed the formation of
multiple systems.  These studies approach the subject from two
distinct perspectives: (1) stellar dynamical models, in which
gravitational interactions are computed between point-mass stars, and
(2) hydrodynamical simulations of protocluster evolution.

Purely dynamical interactions between single stars are clearly
ineffective in producing a sufficiently high fraction of binaries
\citep{1971Ap&SS..13..324A,2004RMxAC..21..156A}.  In addition, binary
orbital periods tend to be too short \citep{C1995}.  Inclusion of
hydrodynamical effects can boost the formation rate of binaries, but
at the cost of reducing further their orbital periods
\citep{2003MNRAS.342..926D,GK2005}.  Increasing the number of stars in
the simulations worsens the problem, in the sense that typically only
a few binaries are formed per star cluster, and higher-order
multiples are very rare \citep{HA1992}.  Simulations which start with
a large proportion of binaries can reproduce the observed binary
frequency \citep{KR1995c,2004RMxAC..21..156A}, but studies of the
formation of higher-order multiples in star clusters containing
primordial binaries but no primordial triples fail to reproduce either
the fraction or the orbital characteristics of triples observed in the
field \citep{PH2004}.

Protocluster evolution, in which gas coagulates to form stars or
subclusters in the form of hierarchical stellar systems, may be able
to account for a high proportion of binary and possibly triple systems
in young star clusters \citep{2003MNRAS.339..577B}.  The
hydrodynamical breakup of protostellar cores could be a dominant
mechanism for the formation of binaries, and possibly triples, in this
environment. These methods, however, fail to produce higher-order
multiplicities \citep{2001ApJ...556..265W}.  From these studies it
would seem that star clusters can form from their parent molecular
cloud with a rich population of binaries and triples, but without
significant numbers of quadruples, quintuples, etc.  In that case,
these multiples must form during the early dynamical evolution of the
cluster, by gravitational interactions between single stars, binaries
and triples.

In this paper we study the formation and reaction rates of multiple
systems in clusters initially containing $\sim100$ stars, including a
sizable fraction of binaries (up to 18\%) and triples (up to 32\%).
The simulations are carried out by directly integrating Newton's
equations of motion to an age of approximately 55\,Myr, corresponding
roughly to the moment of dissolution of these clusters.  During the
evolution, cluster members engage in long-lasting hierarchical
interactions involving up to 11 stars.  We present the dominant
interaction channels for multiple stellar encounters in these
simulations.  The small star clusters with primordial binaries and
triples studied here still under-produce quadruple stellar systems
compared to the observed fractions of the Galactic disc, although the
proportions of quadruples, quintuples and sextuples relative to one
another are roughly consistent with observations.
\section{Initial Conditions}\label{sec:InitialConditions}
The clusters simulated in this study are initialised by selecting
positions and velocities for 100 objects distributed according to a
King \citeyearpar{KI1965,KI1966} model with $W_0 = 6$.  For each
object we randomly draw a mass between $m_{\rm min} = 0.1$M$_{\odot}$
and $m_{\rm max} = 30$M$_{\odot}$ from a Kroupa mass function
\citep{KR1998}, resulting in a total cluster mass of about 45\,\msun.
After adding binary and triple components, the masses of the simulated
clusters average about 67\,\msun.  Finally, we scale the velocities of
all stars and multiple centers of mass in the cluster in such a way
that the entire system is in virial equilibrium, in standard \nbody
units \citep{HM1986}, and scale the cluster to a virial radius $r_{\rm
vir} = 0.1$\,pc, corresponding to a half-mass radius of about
0.08\,pc.  Given the rapid expansion of the clusters in the first few
Myr we consider this a reasonable choice.  We call this model S (see
Tab.\,\ref{tab:summary}). For a $\sim 67$\,\msun\, star cluster in the
solar neighbourhood, the Jacobi radius in the Galactic tidal field is
$r_J \simeq 5.5$\,pc, which is an order of magnitude larger than the
initial tidal radius of the adopted King model.  For clarity we
therefore neglect the tidal field in our simulations.  Note that later
we describe additional simulations with larger initial cluster radii,
in which case this assumption may break down.

Starting from model S, we generate model B by randomly selecting 50
stars, which are converted into binaries by adding a companion
(secondary) star and orbital parameters.  Again, before starting the
simulation, the velocities of all single stars and binary centres of
mass are rescaled so that the entire system is again in virial
equilibrium.  The mass of the secondary is randomly chosen from a
uniform distribution between $m_{\rm min}$ and the mass of the
primary.  The orbital parameters are chosen from empirical
distributions describing approximately the inner orbits of
hierarchical triple systems in the MSC \citep{MSC1997, MSC1999}.  The
orbital period $P_{\rm bin}$ of the binaries is selected by generating
a random variable $X$ between 0 and 1 and setting
\begin{equation}
	P_{\rm in}~(X) = 0.09687 X^{-1} - 0.09677\,{\rm~years}.
\label{Eq:Pbin}\end{equation}
The eccentricity for each binary orbit is generated by selecting another
random number $Y$ uniformly in [0,1), and defining
\begin{equation}
	e_{\rm in}~(Y) = 1.16\,\max(0,\, Y - 0.4).
\label{Eq:Ebin}\end{equation}
If for any binary, the separation at periastron is smaller than five
times the maximum stellar radius (from \cite{EFT1989}), new orbital
parameters are selected randomly.

\begin{figure}
\includegraphics{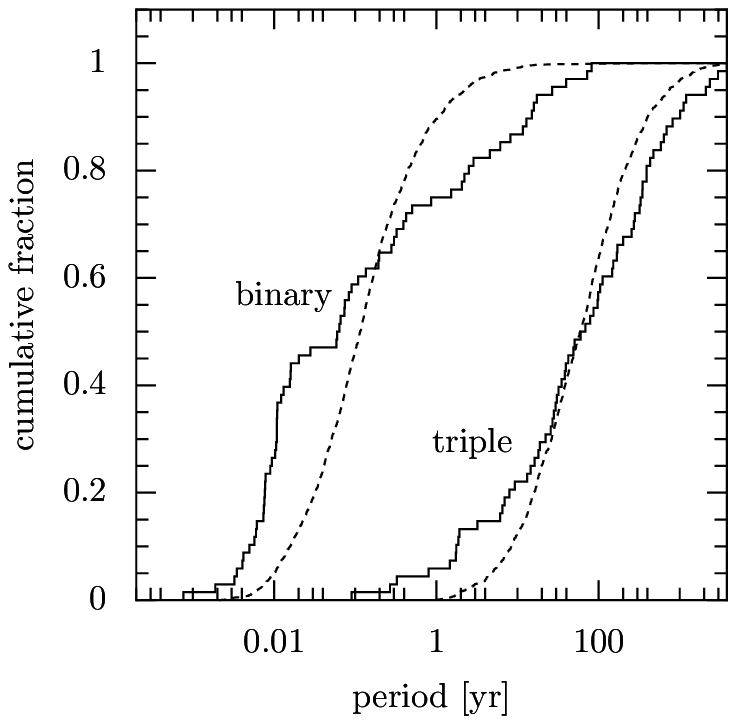}
\caption{Cumulative distribution of the inner and outer orbital
periods of 68 observed triple systems (solid curves; \cite{MSC1997,
MSC1999}), and the distributions (dashed lines) generated from
Eqs.\,\ref{Eq:Pbin} and \ref{Eq:Ptrip}.  }
\label{fig:periods}
\end{figure}

\begin{figure}
\includegraphics{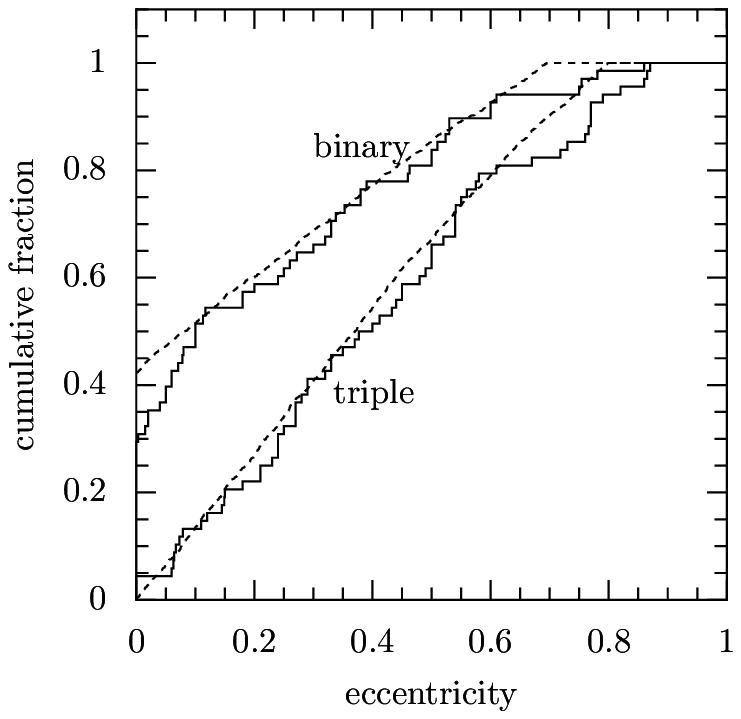}
\caption{Cumulative distribution of the inner and outer orbital
eccentricity for 68 observed triple systems (solid curves;
\cite{MSC1997, MSC1999}), and the distributions (dashed lines)
generated from Eqs.\,\ref{Eq:Ebin} and \ref{Eq:Etrip}.  }
\label{fig:eccentricities}
\end{figure}

These distributions (Eqs.\,\ref{Eq:Pbin} \& \ref{Eq:Ebin}) are
presented in Figures\,\ref{fig:periods} and \,\ref{fig:eccentricities}
and compared with the MSC data.  For clarity, and due to our lack of
understanding of selection effects in the discovery of multiple
systems, we decided deliberately to stay close to the observed
distributions for the orbital period and eccentricity.  An additional
argument for this choice is that our simple prescription for the
generation of initial conditions allows our initial conditions to be
easily reproduced.  The fact that the binary period distribution does
not exactly reproduce the observed distribution is therefore not a
major concern.  The other binary parameters (inclination, orbital
phase and ascending node) are chosen randomly \citep{HB1983}.

Model T is generated by adding a third (outer) star to 32 randomly
selected binaries in model B. The faction of triples thus obtained is
based on the observed fraction of higher order systems in the solar
neighborhood. No simulations were performed with primordial quadruples
or higher-order hierarchies. The mass of the tertiary star is randomly
chosen between $m_{\rm min}$ and the mass of the inner binary.  The
orbital period $P_{\rm trip}$ for the outer orbit of the triple is
selected from a distribution approximating the observed MSC
distribution of outer orbital periods in triples:
\begin{equation}
	P_{\rm out}~(Z) = 46.8 Z^{-1} - 45.8\,{\rm~years},
\label{Eq:Ptrip}\end{equation}
and the eccentricity for the outer orbit is generated with 
\begin{equation}
	e_{\rm out}~(W) = 0.80\,W,
\label{Eq:Etrip}\end{equation}
where $Z$ and $W$ are uniformly distributed between 0 and 1.
The other parameters for the outer orbit are chosen randomly, as for
the binary orbits.  Finally, we check the stability of the triple
using Eq.\,{90} from \cite{MA2001}, and new parameters for the outer
orbit are selected randomly if the triple turns out to be dynamically
unstable, or if the separation at pericentre between the outer star
and the binary is less than five times the maximum stellar radius. As
in model B, the entire system is then restored to virial equilibrium
by rescaling the velocities of single stars and the centre of mass
velocities of binaries and triples.

Figure\,\ref{fig:periods} compares the observed distributions for the
inner and outer orbital period with the distribution generated as
initial conditions for our simulations.
Figure\,\ref{fig:eccentricities} presents a similar comparison for
eccentricities.  Note that our orbital separations and eccentricities
are uncorrelated with one another.
\begin{table}
 \centering
  \begin{tabular}{llllll}
  \hline
   & S & B & T & T$_{1}$ & T$_{3}$\\
	 \cline{2-6}
	$n_\star$   & 100 & 150 & 182 & 182 & 182\\
	$n_{\rm S}$ & 100 &  50 &  50 &  50 &  50\\
	$n_{\rm B}$ &   0 &  50 &  18 &  18 &  18\\
	$n_{\rm T}$ &   0 &   0 &  32 &  32 &  32\\
	$r_{\rm vir}$ [pc] & 0.1 & 0.1 & 0.1 & 1.0 & 3.0\\
	$W_{0}$ & 6 & 6 & 6 & 6 & 6\\
	$N/pc^{3}$ & 50k & 50k & 50k & 50 & 6\\
	%\jelle{removed: $t_{\rm hm}$ [Myr] & $>$ 55 & $>$ 55 & 15 & $>$ 55 & $>$ 55\\}
	$n_{\rm Qd}$ & 0 & 0.5 & 0.78 & 0.25 &0.1\\
	$n_{\rm Qn}$ & 0 & 0.1 & 0.30 & 0.1 &0.01\\
	$n_{\rm Sx}$ & 0 & 0.01 & 0.16 & 0.05 &0.01\\
	$n_{\rm runs}$ & 99 & 77 & 67 & 84 &97\\
\hline
\end{tabular}
\caption{\label{tab:summary} Initial parameters and final fraction of
higher-order multiples for the various simulations.  Each simulation
consists of 100 `particles' which can be single stars, binaries, or
triple systems.  With the adopted number of binaries ($n_{\rm B}$) and
triples ($n_{\rm T}$) the total number of stars $n_\star$ then ranges
between 100 to 182.  Each simulation was initialised from a King model
with $W_0=6$ and with a virial radius $r_{\rm vir} = 0.1$\,pc.  Some
additional simulations were performed with larger virial radii (1\,pc
and 3\,pc).  When the
simulations are terminated at $t=55$\,Myr, each cluster has (on
average) produced $n_{\rm Qd}$ quadruples, $n_{\rm Qn}$ quintuples and
$n_{\rm Sx}$ sextuple systems.  The final row indicates the total
number of simulations performed for each set of initial conditions, each
varying only in the original random seed used.}
\end{table}

\section{Methods}\label{sec:Methods}
Once the initial realizations are generated, our star clusters are
evolved using the {\tt kira} integrator in the Starlab software
package \citep{PM2001}.\footnote{See {\tt
http://www.ids.ias.edu/$\sim$starlab}.}  The equations of motion are
integrated using a fourth-order predictor--corrector Hermite scheme
\citep{MA1992}, using block time steps (\cite{MC1986a,MC1986b,MA1991};
see also \cite{2003Aarseth} for an overview).  Stellar and binary
evolution are modeled with using {\tt SeBa} \citep{PV1996,PH1997}.
For stable hierarchical triples, we follow the internal evolution of
the inner binary, including Roche-lobe overflow, the effect of
stellar-wind mass loss, supernovae and the emission of gravitational
waves.  In our prescription within {\tt SeBa}, the outer star in a
stable triple cannot initiate a phase of mass transfer to the inner
binary by Roche-lobe overflow, but the inner binary is evolved
according to the prescriptions for isolated binaries.  Binaries in
higher-order multiples are evolved, so it is in principle possible to
have a system consisting of six stars, in which three binaries each in
states of mass transfer, orbit one other.  Such a situation, however,
is quite rare.  All simulations were performed on 2.8\,GHz
workstations Intel Pentium 4 processor, and took about 1 hour each.

In order to clarify some of the results on multiple hierarchies we
first provide a brief description of our treatment and identification
procedure for multiple systems.  The {\tt kira} integrator employs a
dynamic tree structure that represents an \nbody\, system as a mainly
`flat' tree having individual stars and the centres of mass of
multiple systems as leaves.  Binary, triple, and more complex
multiples are represented as binary subtrees below their top-level
centre of mass nodes.  The tree structure determines both how node
dynamics is implemented and how the long-range gravitational force is
computed, and also provides a convenient means of identifying
transient structures.

The tree evolves dynamically according to simple heuristic rules:
particles that approach `too close' to one another are combined into a
centre of mass and binary node; and when a node becomes `too large' it
is split into its binary components.  These rules apply at all levels
of the tree, allowing arbitrarily complex systems to be followed.  In
practice, the term `too close' is taken to mean that two objects
approach within roughly the $90^\circ$ deflection distance for typical
stellar masses and speeds (see Figure 7.2 on page 421 of Binney \&
Tremaine 1987, and also Heggie \& Hut 2003, where we adopt $\theta =
\pi$ to compute the $90^\circ$ deflection
distance).\nocite{BT1987,GMBP} `Too large' means that the node's
diameter exceeds $\sim2.5$ times this distance.  The simulation
software prints out summary information each time the tree structure
changes; these data are the basis for the data analysis in this paper.

Reactions between higher order multiple systems are generally chains
of events that cause a particular hierarchy to change, whether or not
the multiple systems are stable.  The start of a reaction is signalled
by a change in the hierarchical structure of a multiple.  In order to
enable a clear identification of specific reactions, we introduce a
time window $\Delta t$, within which a reaction must be completed in
order to be counted.  We set $\Delta t = 0.01$ \nbody time units,
which is about the orbital period of a 1kT binary and which
corresponds to $\sim$550 years in model T.

A reaction can often be decomposed into a series of events in which
only a subset of the total number of stars participates, each
occurring within a smaller time window.  These events are counted as
separate reactions.  Reactions that last for longer than $\Delta t$
are counted multiple times, each time the reaction is counted
according to the dominant hierarchy, which sometimes changes
continuously. Reactions that last for more than $\Delta t$ but that
are do not experience abrupt changes in hierarchy are counted only once.
Thus, short-lived resonances are only counted once, but those which
last for many {\nbody} time units and change configuration may be
counted many times.  Such long-lived resonances, however, are rare.

In our methodology, we distinguish between multiples in terms of the
time they remain intact (as just described).  The survival time of a
multiple is taken to be the interval between the instant when the
multiple forms and the time at which components are lost or new ones
acquired.  If the survival time is more than one \nbody time unit
($1/2\sqrt{2}$ standard crossing times), we call the multiple
`persistent.'  Should the survival time be less than this, the
multiple is called `transient.'  An overview of number of transients
and persistent multiples created in our simulation model T is
presented in tables\,\ref{tab:multipletypes} and
\ref{tab:multiplereduced} in \S\,\ref{sec:mulhierarchy}.
%

%\section{Results}
%
\subsection{Evolution of structural parameters}
\begin{figure}
\includegraphics{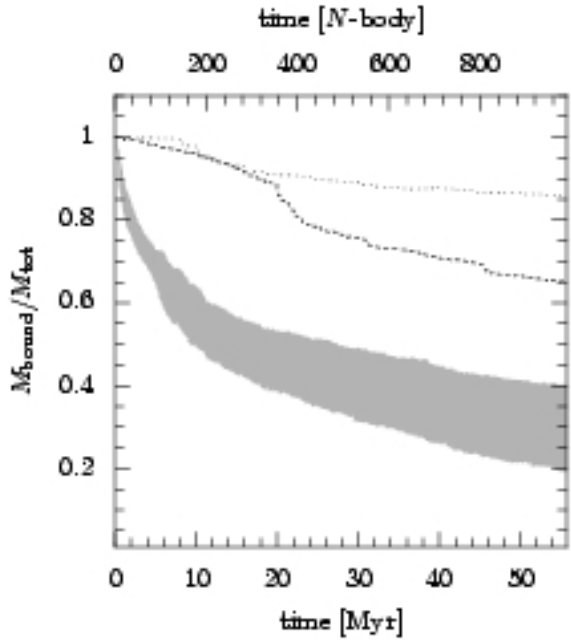}
\caption{Evolution of the bound mass for simulations T (grey area),
$T_1$ (dashed line) and $T_3$ (dots).  The grey shaded region gives
the $1 \sigma$ deviation from the mean for the bound mass of model T.
For the other two models $T_1$ and $T_3$ we only give the mean bound
mass; the dispersions are comparable.  For model T, the upper x-axis
shows time measured in $N$-body units.}
\label{fig:bound_mass}
\end{figure}

We concentrate here on the three models S, B and T with an initial
virial radius $r_{\rm vir} = 0.1$\,pc, but additional simulations have
been performed with $r_{\rm vir} = 1$\,pc and $r_{\rm vir} = 3$\,pc.
The latter two models with primordial triples are identified as T$_1$
and T$_3$ for $r_{\rm vir} = 1$\,pc and $r_{\rm vir} = 3$\,pc,
respectively.  Figure\,\ref{fig:bound_mass} compares the evolution of
the bound mass of models T, T$_1$ and T$_3$.  The means of the T$_1$
and T$_3$ runs are plotted as dashed and dotted curves, respectively.
The gray shaded area in Figure\,\ref{fig:bound_mass} shows the results
(mean $\pm1\sigma$) for the 67 simulations of model T.
Table\,\ref{tab:summary} summarises the results of the various
simulations.

The time at which model T loses half its mass is $t_{\rm hm} =
15^{+15}_{-7}$\,Myr, much shorter than for the other two models.  A
larger initial cluster radius results in a much lower rate of mass
loss.  Mass loss for model T$_3$ is governed by stellar evolution.  In
model $T$, stellar mass loss is less important, and most mass loss is
a result of dynamical ejection.  The sudden divergence between the
curves for models $T_1$ and $T_3$ around 20 Myr is associated with the
onset of significant dynamical activity in $T_1$ at that time (roughly
2--3 initial relaxation times into the evolution).  Models S, B, and T
had comparable overall mass evolution, indicating that an initially
high proportion of triple systems has little influence on the mass
loss rate for these clusters; rather, it is mostly the relaxation of
the cluster that drives dynamical mass loss.  Simulations that
generated more long lived high-order multiples ($n>4$) did, however,
tend to have considerably higher mass loss rates.

\begin{figure}
\includegraphics{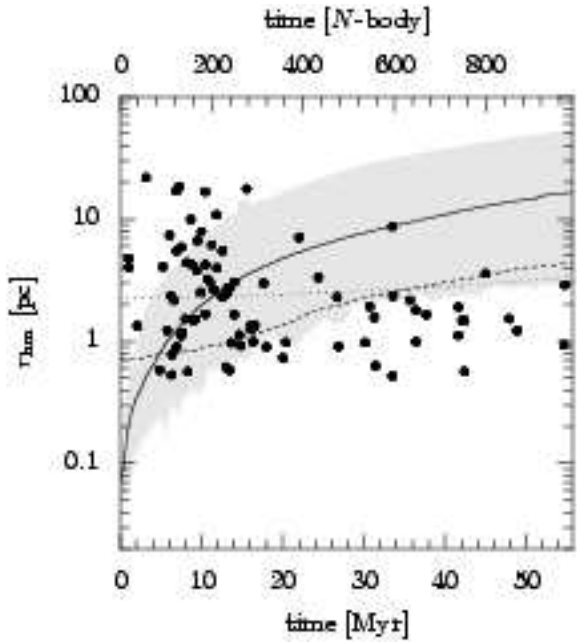}
\caption{Evolution of the half-mass radius for simulations T (grey
area), T$_1$ (dashed line) and T$_3$ (dots).  The grey shaded area
indicates the extrema (from 67 simulations) measured in our
simulations of model T.  For the other two models T$_1$ and T$_3$, we
show only the mean bound mass.  The dispersions for these models are
about 0.5\,pc for model T$_1$ and 1\,pc for model T$_3$.  The filled
circles show the observed star clusters from the open cluster
catalogue \citep{DI2002}.  For model T, the upper x-axis shows time
measured in \nbody units.}
\label{fig:rhalf_real_clusters}
\end{figure}

The evolution of the cluster half-mass radii for model T are presented
in Figure \ref{fig:rhalf_real_clusters}.  Our initial choice of
0.1\,pc as the initial cluster virial radius seems somewhat small
compared with the observations in the open cluster catalogue
\citep{DI2002}.  However, this is quickly compensated by the rapid
expansion of the cluster, driven by a combination of stellar mass
loss, dynamical heating by multiple systems, and mass stratification
\citep{MP2004}.  After about 25\,Myr the simulated cluster has
expanded beyond about 6\,pc, which, according to our earlier estimate,
would exceed the cluster's Jacobi radius in the Galactic tidal field.
Although we ignore the tidal field in our simulations, we argue that a
cluster which expands beyond this radius should be considered
dissolved.  Some of the smaller observed clusters younger than $\sim
20$\,Myr may be satisfactorily reproduced by model T.  At later
times, model T tends to expand too rapidly compared to the observed
cluster population. However, since our models are unlikely to survive
this long if the Galactic tidal field was taken properly into account
the comparison may not be appropriate.  We note, however, that the
expected effect of the Galactic field will be to cause the mass, and
hence the half--mass radius, of the cluster to begin to decline after
25\,Myr, possibly improving the agreement between the model and the
observations, but we do not pursue that possibility further here.  The
initially larger models (T$_1$ and T$_3$) may provide somewhat better
descriptions of the observed clusters at later times.
%% but for these the tidal disruption argument is even more
%% severe.  It seems, from our simulations however, that we have
%% difficulty explaining the older ($\apgt 20$\,Myr) population of star
%% clusters with small radii ($\rm \aplt 2$\,pc) with our adopted initial
%% conditions.

%% There seems to be a marked drop in the observed number of relatively
%% large clusters $r_{\rm hm} \ga 2$\,pc with an age greater than about
%% 15\,Myr.  Interestingly this age corresponds to the time scale on
%% which our simulations of model T lose about half their mass.  By that
%% time the simulation model has expanded by more than a factor of 20,
%% making such clusters very hard to identify, and also prone to
%% stripping by the external tidal field, which was ignored in our
%% simulations.  Such conglomerates in nature, even if bound, will be
%% hard to recognize as clusters \cite{astro-ph/0606749}.

\subsection{Evolution of multiplicity}\label{sec:multiplicity}
\begin{figure*}
\includegraphics{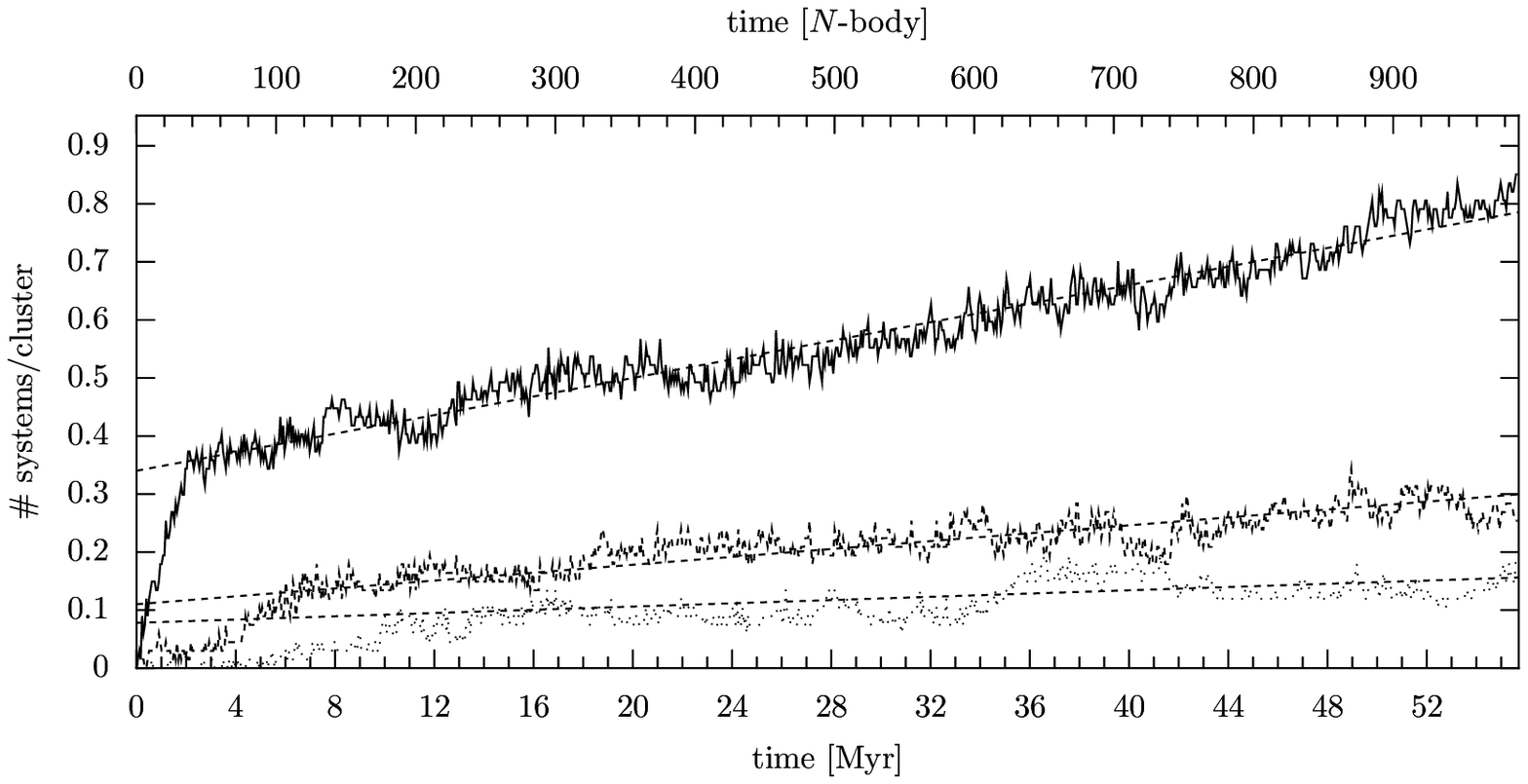}
\caption{Numbers of quadruple (solid curve), quintuple (dashes) and
sextuple (dots) systems per cluster as functions of time. The numbers
plotted are valid for the number of systems present at the end of one
\nbody time unit.  These data are averaged over the 67\,runs of model
T.  The dashed lines are fits through the rightmost portion of the
data (see Eq\,\ref{eq:nsx}).}
\label{fig:quad_quin_nrs}
\end{figure*}

\begin{figure*}
\includegraphics{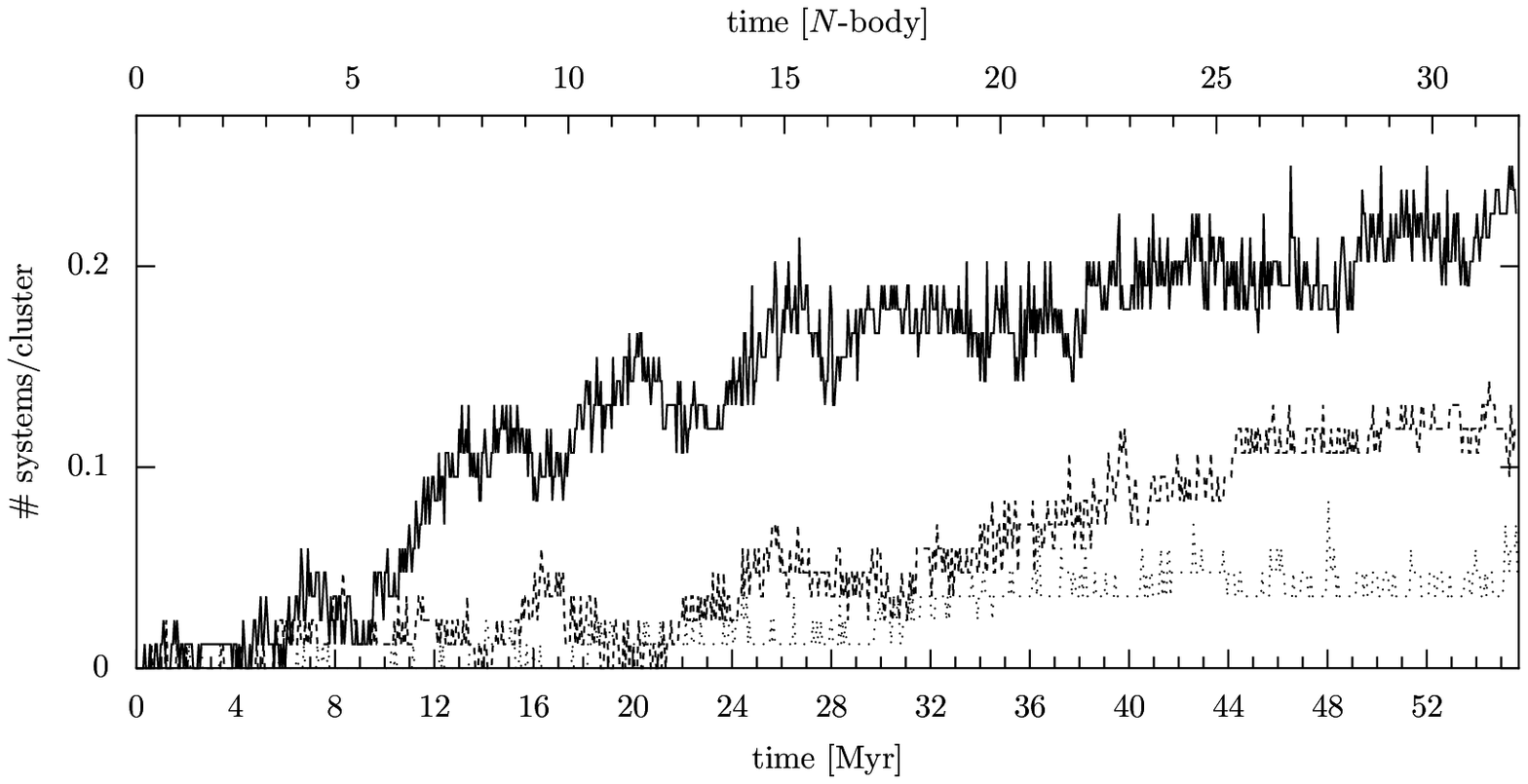}
\caption{Numbers of quadruple (solid curve), quintuple (dashes) and
sextuple (dots) systems per cluster as functions of time. The numbers plotted
are valid for the number of systems present at the end of one \nbody time unit.
These data are averaged over the 84\,runs of model T$_1$.}
\label{fig:quad_quin_nrs_1}
\end{figure*}

Our main simulations (model T) started with only single stars,
binaries and triples, but in time a relatively rich population of
higher-order multiples formed.  During the evolution of model T, the
binary and triple fractions hardly change, while the numbers of stable
hierarchical systems consisting of 4 stars ($n_{\rm Qd}$), 5 stars
($n_{\rm Qn}$) and 6 stars ($n_{\rm Sx}$) increase gradually with
time.  This is illustrated in Figures\,\ref{fig:quad_quin_nrs} and
\ref{fig:quad_quin_nrs_1}, which show the numbers of higher-order
multiples (quadruples, quintuples and sextuples) per cluster as
functions of time for models T and T$_1$.

During the first few million years of model T, the number of
quadruples increases sharply from zero at birth to about 0.35 per
cluster at $t \simeq 2$\,Myr.  After this initial rapid increase, the
formation rate drops by about a factor of 20, and subsequently remains
constant until the end of the simulation.  The numbers of quintuples
and sextuples show a similar trend of rapid increase at early times
and significantly slower growth for the rest of the evolution of the
cluster, but their rates start to increase only after delay times of
about 4 and 10 Myr for the quintuples and sextuples, respectively.
The numbers of quadruples, quintuples and sextuples at late times in
model T are quite well approximated by a simple least-squares linear
fit:
\begin{eqnarray}
    n_{\rm Qd}(t) &=& 8.0 \times 10^{-3} t + 0.34\label{eq:nqd},\nonumber\\
    n_{\rm Qn}(t) &=& 3.4 \times 10^{-3} t + 0.11\label{eq:nqn},\nonumber\\
    n_{\rm Sx}(t) &=& 1.4 \times 10^{-3} t + 0.08
\label{eq:nsx}.
\end{eqnarray}
The formal error on these fits is 4--7\%.  The formation rate drops by
about a factor of two for each higher hierarchy from quadruples to
sextuples. In figure\,\ref{fig:quad_quin_nrs_1} similar trends are
observable for model T$_1$.

It is tempting to interpret the delay in the formation of quintuplets
and sextuples as the time required to establish the channels through
which these higher-order systems form---we first create quadruples
from triples, then quintuples from quadruples, and finally sextuples
from quintuples, and each stage must wait for a reservoir of
lower-order systems to form.  Unfortunately, as demonstrated in
\S\ref{sec:mulreacnet}, addition of a single star is {\em not} the
principal way in which quintuples and sextuples form.  Rather, the
most probable formation channels involve triple--binary and
triple--triple interactions.

Thus, while it is admittedly difficult to discern the relevant
dynamical details in a single set of runs, we interpret
Figure\,\ref{fig:quad_quin_nrs} as follows.  The initial half-mass
relaxation time of model T is $\sim250$ kyr.  The relatively rapid
initial rise in the number of multiples represents the early period of
strong dynamical activity when binaries and triples sink to the
cluster centre and interact, releasing energy and, as a side effect,
producing some higher-order systems.  By 3 Myr the (average) cluster
has expanded fivefold in radius and its density has dropped, reducing
the quadruple formation rate mainly by depleting the numbers of single
stars.  Subsequently, the remnant of the cluster is rich in binaries
and triples and, as the density declines, conditions become more
suitable for forming and retaining quintuple and sextuple systems. (A
typical triple in these runs is hard, in the stellar dynamical sense,
but barely so, making it unlikely that a hierarchical system of two
such triples could form and survive at the initial density of model
T.)

By the end of the simulations (at $\sim 55$\,Myr) a total of 22\% (15
out of 67) of the runs have not produced any persistent hierarchy
consisting of 4 or more stars that survived to the end of the
calculation.  A total of 37\% (25) of the simulations have 1 multiple
present after 55 Myr, 30\% (20) have 2 multiples, and only 10\% (7) of
the runs have 3 stable higher-order multiple systems present after
this time.  None of the simulations had more than three multiples
containing 4 or more stars upon termination.  Short-lived higher-order
systems, however, are common in each of the simulations, and a total
of 7921 higher-order multiples were formed.  However, most were
destroyed before the end of the simulation (see
\S\,\ref{sec:mulhierarchy} for details, and in particular
Table\,\ref{tab:multipletypes}).

Once the cluster disperses, the surviving multiples become part of the
Galactic field, and we may usefully compare their numbers with the MSC
(\S1).  Assuming that the field was assembled from the remnants of
clusters similar to simulation T (of which, on average, $\sim100$
single stars, binaries, and stable multiples remain by the end of the
calculation), the resultant frequencies of triples, quadruples,
quintuples, and sextuples are, respectively, 0.25, 0.008, 0.003, and
0.002.  Thus our particular choice of initial binary and triple
parameters under produces quadruples and higher-order multiples by a
factor of $\sim2$--3 compared to the MSC, although the relative
frequencies of these systems are somewhat encouraging.  We now discuss in
more detail the breakdown of these frequencies within each class.
Figure\,\ref{fig:quad_quin_nrs_1} gives the the equivalent to
Figure\,\ref{fig:quad_quin_nrs} but for simulation $T_1$.  Also in
this case the number of multiples increases quite dramatically in the
first few Myr to level off at later time, although the overall
formation rate of higher order multiples is lower than in the models
$T$. Several higher order multiples were formed in simulation $T_3$,
but their number was rather small and they are omitted from the
figure.

%
%trying to place caption to the right of table
\begin{table}
 \centering
  \begin{tabular}{lllllll}
  \hline
   \small quadruple conf. & \small \#MSC & \small \# total &
	  \small $\overline{\rm dt}$ & \small \# pers.&\small
	   $\overline{\rm dt}$ \\
 \hline
\tiny (B,B)     & 69 & 1171 & 5.080 & 290 & 20.1\\
\tiny ((B,S),S) & 61 & 1093 & 2.43 & 209 & 12.1\\
\tiny ((S,B),S) & 4  & 945 & 0.97 & 165 & 4.66\\
\tiny (S,(B,S)) & 4  & 358 & 4.45 & 41 & 38\\
\tiny (S,(S,B)) & 0  & 297 & 0.62 & 37 & 4.0\\
\cline{2-2}\cline{3-3}\cline{5-5}
 total          &138 & 3864& & 742 &\\
\hline
  \small quintuple conf. & \small \#MSC & \small \# total & \small 
	$\overline{\rm dt}$ & \# pers.&\small $\overline{\rm dt}$ \\
\hline
\tiny ((B,S),B) & 9 & 949 & 3.08 & 202 &13.9\\
\tiny (B,(B,S)) & 7 & 854 & 2.16 & 145 &11.9\\
\tiny ((B,B),S) & 5 & 142 & 2.21& 20 &15\\
\tiny (((B,S),S),S) & 4 & 14 & 0.33 & 2 & 2\\
\tiny (B,(S,B)) & 0 & 205 & 0.83 & 32 &4.5\\
\tiny ((S,B),B) & 0 & 162 & 1.83 & 29 &9.4\\
\tiny (S,(B,B)) & 0 & 112 & 0.70 & 23 &2.8\\
\tiny (((S,B),S),S) & 0 & 12 & 0.71 & 3 & 3\\
\tiny (S,(S,(B,S))) & 0 & 10 & 0.43 & 2 &2\\
\tiny ((S,(B,S)),S) & 0 & 12 & 0.13 & 1 & 1\\
\tiny (S,((S,B),S)) & 0 & 17 & 0.07 & 0 &0\\
\tiny (S,((B,S),S)) & 0 & 5 & 0.01 & 0 &0\\
\tiny ((S,(S,B)),S) & 0 & 4 & $<$ 0.01 & 0 &0\\
\tiny (S,(S,(S,B))) & 0 & 3 & $<$ 0.01 & 0 &0\\
\cline{2-2}\cline{3-3}\cline{5-5}
 total        &25& 2501 & & 460 &\\
\hline
  \small sextuple conf. & \small \#MSC & \small \# total & \small
	 $\overline{\rm dt}$ & \small \# pers.&\small $\overline{\rm
	   dt}$ \\
\hline
\tiny ((B,B),B) & 3 & 88 & 6.0 & 28 & 19\\
\tiny (((S,B),S),B) & 2 & 1 & 0.01 & 0 & 0\\
\tiny ((B,S),(B,S)) & 1 & 699 & 1.74 & 132 & 8.53\\
\tiny ((B,S),(S,B)) & 1 & 121 & 0.74 & 11 & 6.7\\
\tiny ((S,B),(B,S)) & 0 & 381 & 0.36 & 15 & 7.6\\
\tiny (B,(B,B)) & 0 & 67 & 11 & 11 & 66\\
\tiny (((B,S),B),S) & 0 & 46 & 0.33 & 4 & 3\\
\tiny ((S,B),(S,B)) & 0 & 21 & 1.2 & 1 & 30\\
\tiny (S,((B,S),B)) & 0 & 21 & 0.38 & 1 & 4\\
\tiny ((B,(B,S)),S) & 0 & 17 & 13 & 3 & 71\\
\tiny (B,((B,S),S)) & 0 & 14 & 0.57 & 2 & 3\\
\tiny (((S,B),B),S) & 0 & 13 & 3.7 & 0 & 0\\
\tiny (S,((S,B),B)) & 0 & 9 & 0.6 & 2 & 2\\
\tiny (S,(B,(B,S))) & 0 & 8 & 2 & 2& 9\\
\tiny (B,(S,(B,S))) & 0 & 8 & 1 & 2 & 4\\
\tiny (((B,B),S),S) & 0 & 6 & 0.4 & 1 & 2\\
\tiny (S,((B,B),S)) & 0 & 5 & 0.5 & 1 & 2\\
\tiny (S,(B,(S,B))) & 0 & 5 & 0.3 & 0 & 0\\
\tiny (((B,S),S),B) & 0 & 4 & 0.2 & 0 & 0\\
\tiny ((B,(S,B)),S) & 0 & 4 & 0.08 & 0 & 0\\
\tiny ((S,(B,B)),S) & 0 & 2 & 0.01 & 0 & 0\\
\tiny ((S,(B,S)),B) & 0 & 3 & 0.8 & 2 & 1\\
\tiny ((S,(S,B)),B) & 0 & 1 & 0.05 & 0 & 0\\
\tiny (B,((S,B),S)) & 0 & 3 & $<$ 0.01 & 0 & 0\\
\tiny (B,(S,(S,B))) & 0 & 3 & 0.1 & 0 & 0\\
\tiny (S,(((S,B),S),S)) & 0 & 2 & 10 & 1 & 20\\
\tiny ((((S,B),S),S),S) & 0 & 2 & 0.09 & 0 & 0\\
\tiny (S,(((B,S),S),S)) & 0 & 2 & 0.01 & 0 & 0\\
\tiny (((S,(B,S)),S),S) & 0 & 1 & 0.8 & 0 & 0\\
\cline{2-2}\cline{3-3}\cline{5-5}
total         &7&  1556 & &227 &\\
\hline
\end{tabular}
  \caption{Overview of the stellar multiplicities observed in the MSC
  and in our 67 simulations of model T.
  The first column gives this configuration of the multiple. A single
  star is designated by an S, a binary by a B. A binary consists
  of two single stars and could be written as (S, S) $\equiv$ B.  The
  entries are ordered by mass, with the more massive component always
  positioned on the left.  The parentheses indicate the hierarchy of
  the multiple system.
} \label{tab:multipletypes} 

\end{table}
\begin{table}
\contcaption{
  We encounter two stable configurations for triple systems, (S, B)
  and (B, S), 5 stable configurations for quadruples, 14 for
  quintuples, and 29 for sextuples.  The second column presents the
  stable configurations listed in the MSC.  The next two columns give
  the number of transient and persistent systems in our simulation
  model T, and the mean time (in \nbody\, units) these systems survive
  before engaging in a new reaction ($\overline{\rm dt}$).
  The last two columns give the number of multiples of each
  configuration that survive for at least for one N-body time unit
  (about 55\,kyr).  The final column gives their average lifetimes in
  \nbody time units.  Note that this average omits systems that last
  for the entire duration of the simulation.
  The total numbers of formations per multiple order (quadruple,
  quintuple, etc.) are presented at the bottom of each column.  A
  summary and a comparison with observed systems is presented in
  Table\,\ref{tab:multiplereduced}.}
 \end{table}

\subsection{The hierarchy of multiple systems}\label{sec:mulhierarchy}
During the simulations, the configurations of the multiple systems
continually change, as do the numbers of components.  In this section
we discuss the various hierarchical structures, and the frequencies
with which they appear in the simulations.  In
\S\,\ref{sec:mulreacnet} we further explore the channels through which
higher-order multiples are formed and destroyed.  For both this
subsection and the next we concentrate on model T.

In tables\,\ref{tab:multipletypes} and \ref{tab:multiplereduced} we
present an overview of the transients and persistent multiples created
in simulation model T.  The multiples which escape the cluster
are included in this table, as well as the multiples which remain
bound.  A single star in these tables is identified by the letter S.
We denote a pair of bound objects by putting parentheses around them,
with the more massive of the two always to the left.  For brevity we
introduce separate notation for a binary (a pair of stars), which we
write as B $\equiv$ (S,S).  A triple can have two configurations:
(B,S) if the binary is more massive than the star orbiting it, or
(S,B) if the outer star is more massive.
Table\,\ref{tab:multipletypes} draws a distinction between the total
number of configurations and those which are persistent.
Table\,\ref{tab:multiplereduced} compares the simulated hierarchies
directly with those observed.

The most long-lived multiple configuration, on average, consists of 4
stars arranged in the classical quartet (B,B), in which two binaries
orbit one another.  A total of 1171 reactions led to such a
configuration.  Of these, 290 resulted in persistent systems, with an
average lifetime of about 20\,Myr.  Aarseth
\citeyearpar{2004RMxAC..21..156A} has commented on the formation of such
binary--binary systems, and on their longer lifetimes compared to
other, less compact, configurations.  Due to their smaller cross
sections, they are less likely to have a fatal encounter with another
object than are strictly hierarchical systems.  The ratios of the
numbers of the various quadruple configurations to the total number of
quadruples in the MSC are: 0.50 for (B,B), 0.44 for ((B,S),S), 0.05
for ((S,B),S) and (S,(B,S)). The configuration (S,(S,B)), in which the
inner binary is less massive than the two single stars orbiting it, has
not been observed.

To compare these numbers with the results of our simulations of model
T, we multiply the frequencies at which these various configurations
are formed by their average lifetimes.  The probability of observing
any of the above quadruple configurations during a simulation is then
0.54 for (B,B), 0.23 for ((B,S),S), 0.14 for ((S,B),S), 0.07 for
(S,(B,S)), and 0.01 for the unobserved (S,(S,B)).  These relative
ratios are comparable to those actually observed.

In the observations it is often hard to determine the masses of the
component stars, so we present in Table\,\ref{tab:multiplereduced} a
reduced version of Table\,\ref{tab:multipletypes}, in which we list
the observed numbers of hierarchies, but group all systems with the
same physical hierarchy together.  Thus, quadruple systems are reduced
to just two types (B,B) and ((B,S),S), the latter category then
contains all possible permutations in which a binary is orbited by two
single stars: ((B,S),S), ((S,B),S), (S,(B,S)) and (S,(S,B)).  Among
the quintuple configurations, the strictly hierarchical systems
((B,S),S),S) are vastly outnumbered by the triple/binary pairs
((B,S),B) in the simulations as well as in the observations. In the
observed sample, however, permutations of ((B,S),B) are much more
abundant than in our simulations.  The differences between the
occurrences of observed and simulated multiplicities is quite
striking.  The origin of these discrepancies is not trivial to
understand; it could stem from observational selection effects or be a
result of our choice of initial binary and triple parameters.
\begin{table}
 \centering
  \begin{tabular}{lllllll}
  \hline
   \small configuration & \small \#MSC & \small \# model T \\
 \hline
\multicolumn{3}{l}{Common quadruple configurations} \\
\tiny ((B,S),S)     & 69 & 17.2 $\pm$ 0.8 \\
\tiny (B,B)         & 69 & 20.2 $\pm$ 1.2 \\
\hline
\multicolumn{3}{l}{Common quintuple configurations} \\
\tiny (((B,S),S),S) &  4 &  0\\
\tiny ((B,B),S)     &  5 &  1.1 $\pm$ 0.2 \\
\tiny ((B,S),B)     &  9 & 13.3 $\pm$ 0.6 \\
\hline
\multicolumn{3}{l}{Common sextuple configurations} \\
\tiny ((B,B),B)     & 3  &  4.0 $\pm$ 0.6 \\
\tiny ((B,S),(B,S)) & 1  &  4.2 $\pm$ 0.4 \\
\tiny (((B,S),S),B) & 2  &  0.9 $\pm$ 0.3 \\
\hline
\end{tabular}
\caption{Reduced and renormalised version of
Table\,\ref{tab:multipletypes}.  Here all possible configurations are
summarised without discriminating among possible permutations.  The
number of observed systems is given in the second column.  The final
column shows the number of these configurations expected based on the
persistent systems seen in model T.  These numbers are computed on the
assumption that all stars and multiples in the solar neighbourhood
formed in clusters similar to model T, and are scaled to the
`survivor' frequencies presented in the text
(\S\ref{sec:multiplicity}), assuming a total of 4800 stars and
multiples in the MSC survey.  The quoted error is the Poissonian
uncertainty based on the total number of stable configurations
occurring in simulation T.  }
\label{tab:multiplereduced}
\end{table}

%\subsection{Multiplicity reaction network}\label{sec:mulreacnet}
\subsection{Multiplicity formation and destruction
  reactions}\label{sec:mulreacnet}
A gravitational dynamical interaction in which the multiplicity of one
object changes can be described in terms of creation and destruction
reactions.  Upon the dissociation of a binary, two single stars
appear, and when two single stars form a bound pair a binary is
created.  A cascade of such reactions then becomes a multiplicity
reaction network.  It is straightforward to extend this description to
reactions in which more than two stars participate, although the
reactions and the reaction network can become quite complicated.  We
now use this perspective to describe the formation channels for
systems consisting up to six stars.

%\begin{table}
% \centering
%  \begin{tabular}{lll}
%  \hline
%   \# multiples & fraction of clusters & \# of clusters\\
% \hline
% 0 & 0.22 & 15\\
% 1 & 0.37 & 25\\
% 2 & 0.30 & 20\\
% 3 & 0.10 & 7\\
%\hline
%\end{tabular}
%  \caption{The fractions of clusters with a certain number of multiples of multiplicity larger than 3 in a total of 67 simulations.}
%\label{tab:formedmultiples}
%\end{table}
%
\begin{figure*}
\begin{center}
$\begin{array}{ccc}
\mbox{\emph{creation}} & & \mbox{\emph{destruction}}\\
\begin{minipage}[t]{0.4\textwidth}
\vspace{0pt}
\centering
\includegraphics[scale=0.7]{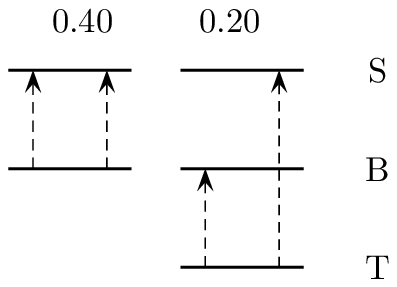}
\end{minipage} & {\hfill} &
\begin{minipage}[t]{0.4\textwidth}
\vspace{0pt}
\centering
\includegraphics[scale=0.7]{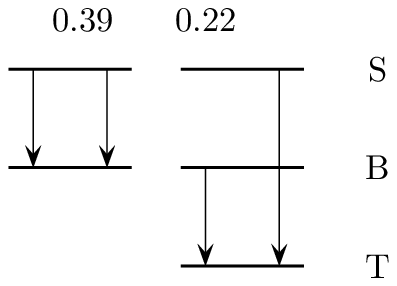}
\end{minipage}
\end{array}$
\end{center}
\caption{Dynamical reactions for the creation (left) and destruction
(right) of single stars.  In 67 simulations of model T, a total of
4330 unique creation reactions were identified, versus 3245 unique
destruction reactions.  Unique meaning that if a particular reaction
involving a certain set of single star components occurred more than
once during a simulation it is counted only once. A reaction results
in the creation or destruction of at least one single star.  In the
leftmost creation reaction, for example, two single stars are created
from the destruction of a single binary via stellar dynamics such as a
supernova event of due to the background potential of the star
cluster, whereas in the second creation reaction one single star and
one binary form from a triple.  The numbers along the top line show
the relative frequency in relation to all reactions encountered of
this particular reaction. Note that only the most common reactions are
given, hence these numbers do not add up to unity. The text to the
right indicates the state of the system.  The following terminology is
used: S for single stars, B for binaries, T for triples, Qd, Qn and Sx
for quadruples, quintuples and sextuples, respectively.  }
\label{fig:singlereactions}
\end{figure*}
\begin{figure*}
\begin{center}
$\begin{array}{ccc}
\mbox{\emph{creation}} & & \mbox{\emph{destruction}}\\
\begin{minipage}[t]{0.4\textwidth}
\vspace{0pt}
\centering
\includegraphics[scale=0.7]{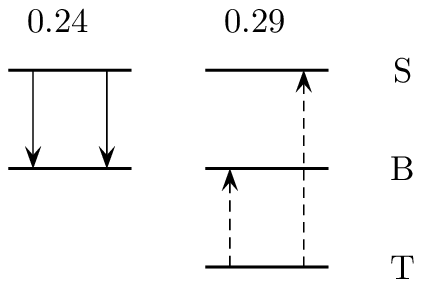}
\end{minipage} & {\hfill} &
\begin{minipage}[t]{0.4\textwidth}
\vspace{0pt}
\centering
\includegraphics[scale=0.7]{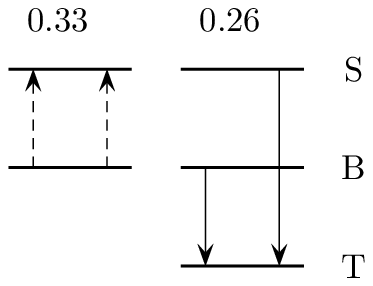}
\end{minipage}
\end{array}$
\end{center}
\caption{Principal reactions leading to the creation and destruction
of binaries: 2639 creation and 2628 destruction reactions.}
\label{fig:binaryreactions}
\end{figure*}
\begin{figure*}
\begin{center}
$\begin{array}{ccc}
\mbox{\emph{creation}} & & \mbox{\emph{destruction}}\\
\begin{minipage}[t]{0.4\textwidth}
\vspace{0pt}
\centering
\includegraphics[scale=0.7]{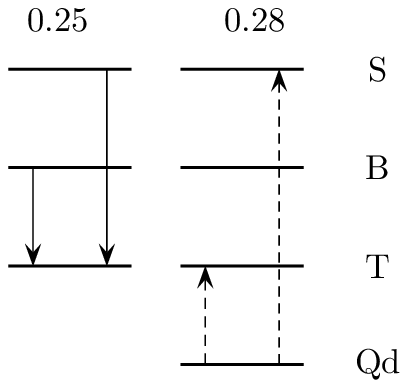}
\end{minipage} & {\hfill} &
\begin{minipage}[t]{0.4\textwidth}
\vspace{0pt}
\centering
\includegraphics[scale=0.7]{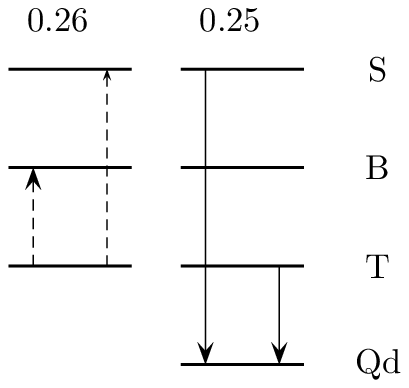}
\end{minipage}
\end{array}$
\end{center}
\caption{Reactions leading to the creation and destruction of triples:
2700 creation and 2969 destruction reactions.}
\label{fig:triplereactions}
\end{figure*}
\begin{figure*}
\begin{center}
$\begin{array}{ccc}
\mbox{\emph{creation}} & & \mbox{\emph{destruction}}\\
\begin{minipage}[t]{0.4\textwidth}
\vspace{0pt}
\centering
\includegraphics[scale=0.7]{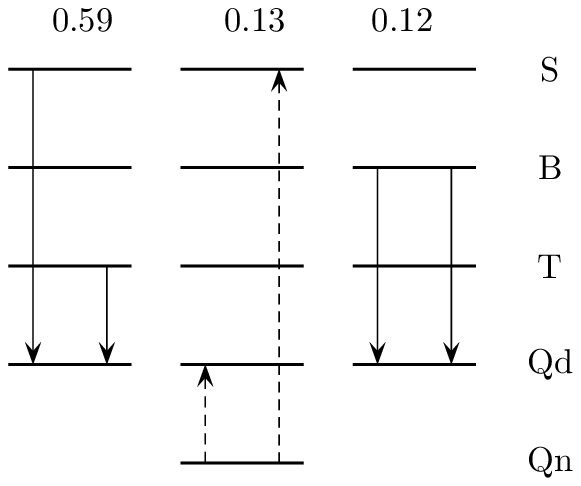}
\end{minipage} & {\hfill} &
\begin{minipage}[t]{0.4\textwidth}
\vspace{0pt}
\centering
\includegraphics[scale=0.7]{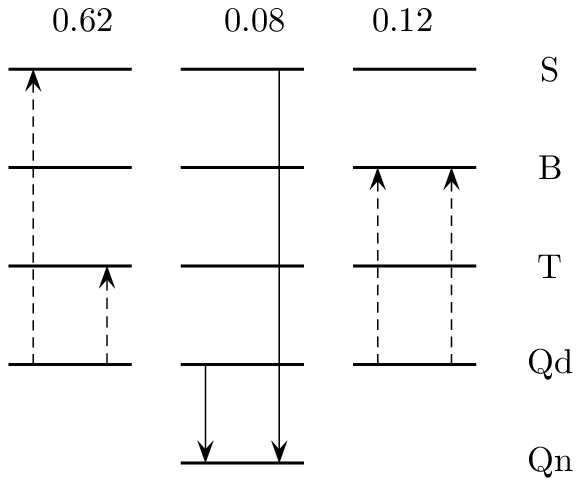}
\end{minipage}
\end{array}$
\end{center}
\caption{Reactions leading to the creation and destruction of
quadruples: 1247 creation and 1229 destruction reactions.}
\label{fig:quadruplereactions}
\end{figure*}
\begin{figure*}
\begin{center}
$\begin{array}{ccc}
\mbox{\emph{creation}} & & \mbox{\emph{destruction}}\\
\begin{minipage}[t]{0.4\textwidth}
\vspace{0pt}
\centering
\includegraphics[scale=0.7]{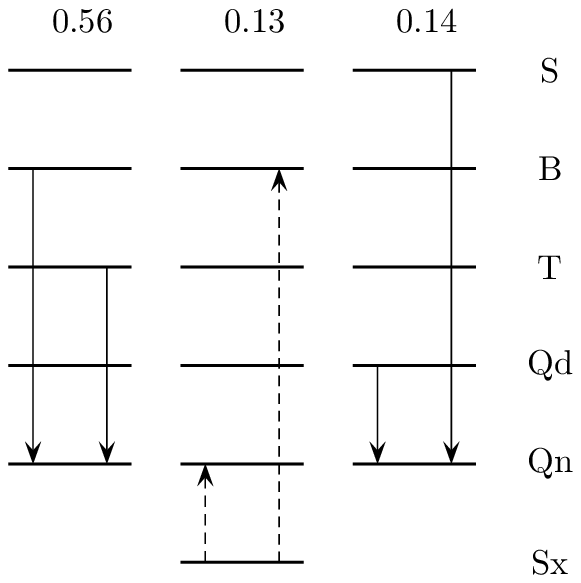}
\end{minipage} & {\hfill} &
\begin{minipage}[t]{0.4\textwidth}
\vspace{0pt}
\centering
\includegraphics[scale=0.7]{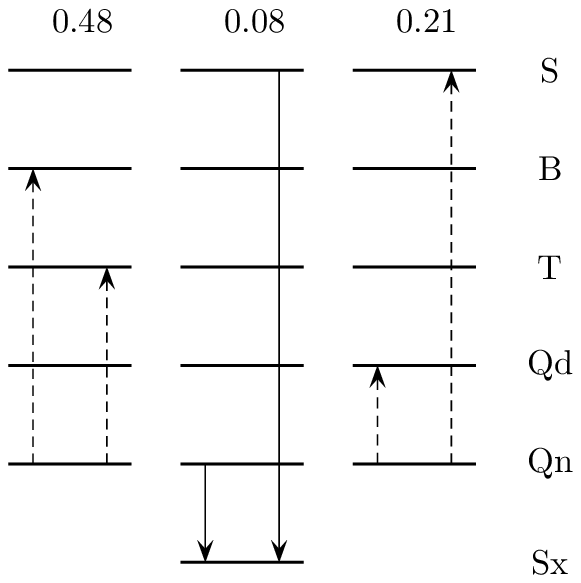}
\end{minipage}
\end{array}$
\end{center}
\caption{Reactions leading to the creation and destruction of
quintuples: 724 creation and 760 destruction reactions.}
\label{fig:quintuplereactions}
\end{figure*}
\begin{figure*}
\begin{center}
$\begin{array}{ccc}
\mbox{\emph{creation}} & & \mbox{\emph{destruction}}\\
\begin{minipage}[t]{0.4\textwidth}
\vspace{0pt}
\centering
\includegraphics[scale=0.7]{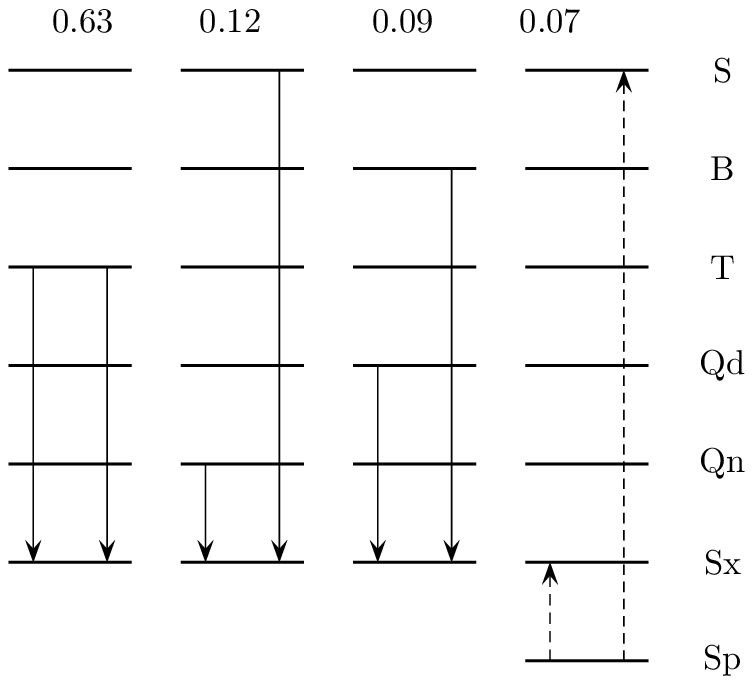}
\end{minipage} & {\hfill} &
\begin{minipage}[t]{0.4\textwidth}
\vspace{0pt}
\centering
\includegraphics[scale=0.7]{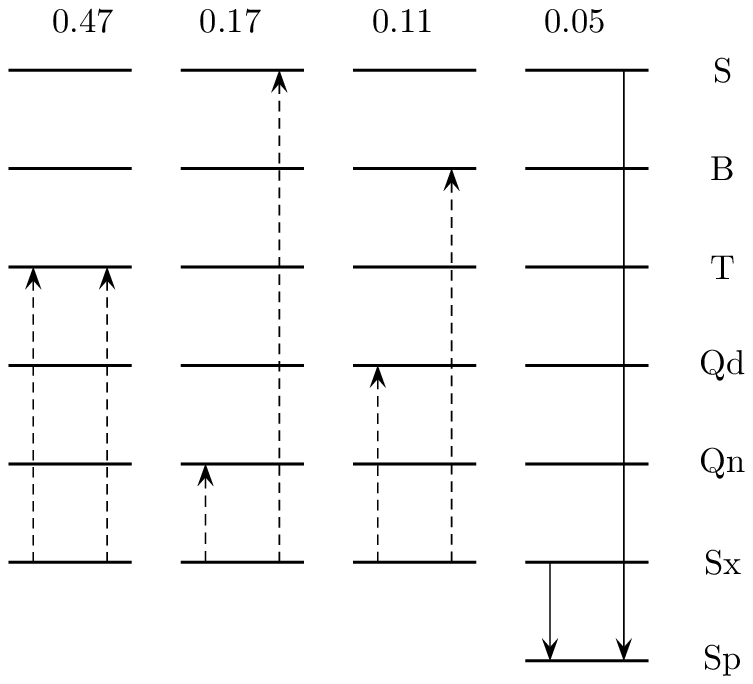}
\end{minipage}
\end{array}$
\end{center}
\caption{Reactions leading to the creation and destruction of
sextuples: 511 creation and 530 destruction reactions.}
\label{fig:sextuplereactions}
\end{figure*}

The configurations listed in Table\,\ref{tab:multipletypes} and
Table\,\ref{tab:multiplereduced} are often formed in complex
interactions involving single stars and higher-order systems.  In this
section we concentrate on the individual formation and destruction
reactions for multiplicities of up to 6 stars.

Figure\,\ref{fig:singlereactions} presents the most common reactions
leading to the `creation' (liberation) of single stars due to the
gravitational decay of multiple systems (left), and the corresponding
`destruction' (consumption) reactions resulting in the loss of single
stars (right).  Each `ladder' represents a reaction channel; the
symbols S, B and T on the rungs refer to single, binary and triple
stars, respectively.  An arrow is drawn from the originating level to
the final state of the system.  Dashed arrows indicate the reduction
of a higher-order system to one of lower order---a binary being
disrupted or a triple being reduced to a single star and a binary.
Such a reduction could be caused by an encounter with another cluster
member, by stellar evolution (e.g. a supernova event) or, in the case
of triple reduction, by the internal dynamics of a dynamically
unstable system.  These different scenarios are not represented
separately in these diagrams because the effect on the multiplicity of
the components is the same.  Solid arrows indicate capture, as when
a single star or small system combines (typically in the presence of
another cluster member) with another small system to form a higher
order multiple.  Note that, although this presentation might suggest
that interactions occur in isolation, this is usually not the case.
Many interactions take place near the core of the cluster, and there
may be many other stars nearby which can carry away small amounts of
energy and angular momentum.  It is sometimes hard to identify the
star(s) responsible for the binding energy and angular momentum of a
post-encounter bound pair.  Our simulations include stellar and binary
evolution and allow for stellar collisions to take place, but in order
not to unnecessarily complicate matters, these processes are not
displayed separately here.

The number at the top of each reaction channel indicates its relative
frequency in our simulations.  Thus the left panel in
Figure\,\ref{fig:singlereactions} presents reactions in which single
stars are created by the ionisation of a binary (40\%) or by
destruction of a hierarchical triple (20\%), whereas the right panel
describes the `destruction' of single stars by absorption into a
higher-order system binary (39\%) or triple (22\%).  The sums of the
creation and destruction frequencies should each be unity, but for
clarity only the most important reactions are shown.  The remaining
interactions comprise a wealth of low-probability, sometimes rather
arcane reactions contributing to the overall cluster evolution.

Figures\,\ref{fig:binaryreactions}\,--\,\ref{fig:sextuplereactions}
show analogous reaction channels for the formation and destruction of
binary through sextuple systems.  The symbols Qd, Qn and Sx refer to
quadruples, quintuples and sextuples.  Note the important role played
by triples in the formation of higher-order multiples.  Quadruples and
quintuples are most commonly formed from a triple that absorbs a
single star or a binary, whereas sextuples tend to form from the
combination of two triples.  (See also Table\,\ref{tab:multipletypes},
where sextuples consisting of two triples orbiting one another are
relatively common.)  We find similar trends in the destruction of the
higher-order multiples.  In particular, the decay of sextuples to form
two triples is quite striking.

The reaction frequencies presented in these figures are computed for
the entire simulation of model T, even though one might argue that the
early phase of rapid multiple creation ($t < 2$\,Myr) could produce
reactions different from those found the later, slower phase of the
evolution (Fig. \ref{fig:quad_quin_nrs}).  Interestingly, there is no
significant difference between the types of reactions seen during the
early and late phases.

\subsection{The orbital parameters of the hierarchies}\label{sec:orbits}

It is not trivial to compare orbital parameters of the observed
multiples with those in the simulations, in part due to the large
number of parameters and due to severe selection effects.  Complete
orbital solutions are not available for any of the multiples listed in
the MSC, but for four quadruples the MSC lists the orbital periods and
eccentricities for each of the three orbits and also the masses of the
four stars; those systems are HD\,08065+1757, HD\,05569+0939,
HD\,11128+3205 and HD\,11171-2414.

\begin{figure}
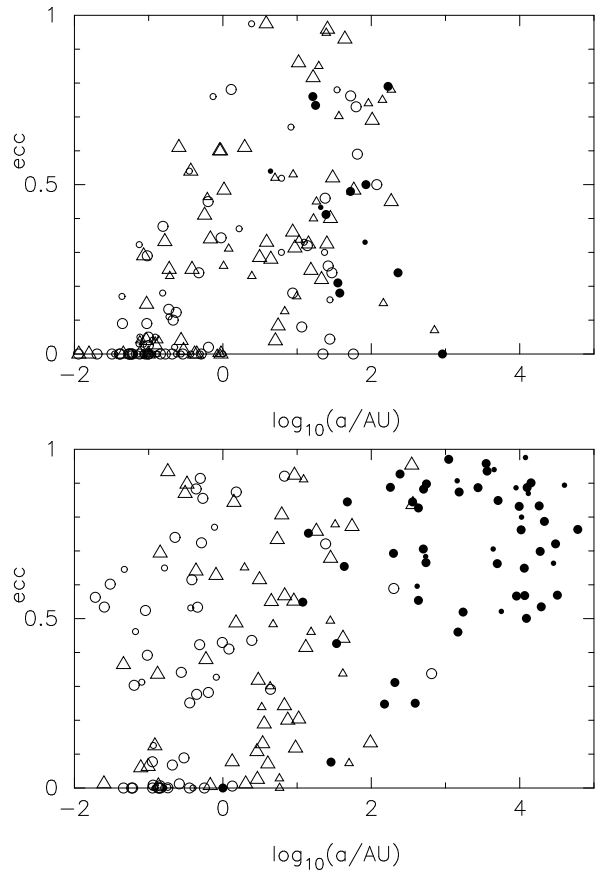

\begin{center}
\includegraphics[angle=-90,scale=0.35]{fig_ae_MSC4ds.ps}
\includegraphics[angle=-90,scale=0.35]{fig_ae_Sim4ds.ps}
\end{center}
\caption{Semi major axis and eccentricity of quadruples observed in
the MSC (top panel) and those considered persistent by the end of
simulation T (bottom panel).  Each quadruple is identified by three
sets of orbital parameters; however, for the MSC, some parameters are
often unknown.  The small symbols represent hierarchical quadruples
((B, S), S), whereas the large symbols are for the (B, B)
configurations.
For the hierarchical systems ((B, S), S), bullets indicate the orbital
parameters for the outermost orbit, circles represent the inner
binary, and triangles indicate orbital parameters for the intermediate
star orbiting the inner binary.
For the (B, B) configuration, bullets indicate the relative orbital
parameters for the two binaries, triangles indicate orbital parameters
for the binaries containing the most massive stars, and triangles
represent the less massive binary.
Note that for the observed sample the orbital parameters are generally
not all known, whereas for the simulations orbital elements for all
quadruples are plotted.}
\label{fig:Orbital_parameters_quads}
\end{figure}

Nevertheless we attempt to compare the orbits of the multiples in the
MSC with those obtained in our simulations. For clarity we limit
ourselves here to the population of quadruple systems.
Fig.\,\ref{fig:Orbital_parameters_quads} (top panel) shows the orbital
separations and eccentricities for all quadruples in the MSC.  The
bottom panel shows the same information for persistent quadruples at
an age of 55\,Myr for model T.
Small symbols indicate hierarchical systems ((B, S), S), whereas large
symbols show data for the (B, B) configuration.  For hierarchical
quadruples, the outermost, intermediate, and inner orbits are
represented by small bullets, triangles, and circles, respectively.  For
binary--binary systems, the large bullets represent the outer orbit,
while the large triangles and circles indicate, respectively, the
inner orbits containing the most massive and least massive stars.
%
%% The bullets give parameters for the outermost orbit---the outer star
%% in the hierarchical case, the binary--binary orbit otherwise.  The
%% small triangles indicate the orbital parameters for the intermediate
%% orbit in a hierarchical system; for a binary--binary system, the large
%% triangles indicate the orbital parameters for the binary containing
%% the most massive star.  Small circles indicate the orbital parameters
%% for the innermost binary in a hierarchical system; large circles give
%% orbital parameters for the less massive binary in a (B, B) system.

Many of the `outer' orbits in the observed quadruples are missing from
the sample in the MSC, which reflects the few bullets in
Fig.\,\ref{fig:Orbital_parameters_quads}. The outer orbits seem to be
cut off at around $10^3$\,AU. The simulated sample of outer orbits,
however, extends all the way to about $10^5$\,AU. The few known outer
orbits in the observed sample of quadruple systems strongly suggests
that these orbits are simply not observed, and that the observational
selection effect becomes important at around a few hundred AU.

There does not seem to be a specific selection effect regarding the
orbital parameters of the inner binary orbits in the 69 (B, B)
quadruples listed in the MSC, as for the primary as well as the
secondary binary a total of 46 are observed. The inter-binary orbit,
however, is quite heavily affected by selection effects, as the MSC
lists only 10 binaries with known period and eccentricity (the large
bullets in the top panel of Fig.\,\ref{fig:Orbital_parameters_quads}).

These different selection effects are best seen in the large number of
missing bullets in the top panel of
Fig.\,\ref{fig:Orbital_parameters_quads}, which suggests that the
outer orbits in hierarchical systems, as well as the inter-binary
orbits in (B, B) systems, are severely affected by observational
selection effects. The other seem to be rather well represented by the
observations, with the possible exception of the larger number of
circularized binaries in the observation and some excess of short
period eccentric binaries in the simulations. This suggests that tidal
circularisation in our simulations is less effective that in nature.
This should not come as a surprise, as tidal effects are not taken
into account in higher order systems in which an inner binary is
perturbed by an outer object, as is often the case in these quadruple
systems.

We have performed similar analyses comparing the orbital parameters
for the quintuples and sextuple systems in the MSC with simulation T,
and the results of these are quite similar, in the sense that the
observational selection effects for determining orbital parameters
gradually increase for larger periods.  In these cases we also find
that the simulations tend to overproduce short-period systems with
high eccentricities relative to the observations, at the expense of
short-period circular orbits. We decided not to show the plots for
these higher order multiples, as the quality of the observational data
declines with increasing multiplicity.

\section{Discussion and Conclusions}\label{sec:discon}
We have performed simulations of star clusters which initially
consisted of single stars, binaries and triples. The initial
conditions were selected to match the young ($< 50$\,Myr) star
clusters in the solar neighbourhood, consisting of at most a few
hundred stars.  The initial conditions for the binaries and triples
were selected based on the observed populations, which we transform
directly to input parameters.  For simplicity, observational selection
effects are neglected, both in determining the distribution functions
from which we generate our initial conditions, and in our final
comparison with observations.

After initialisation we run the simulations for about 55\,Myr
(corresponding to about 1000 N-body time units, or about 350 dynamical
crossing times of the system) and compare our results with the
observed star clusters and multiples in the field.  Our models with
initial half-mass radii of about 0.1\,pc are consistent with several
observed young ($\aplt 20$\,Myr) star clusters.  The rapid expansion
of our simulated clusters is attributed partly to stellar mass loss
and partly to dynamical effects.  Our simulations fail to explain the
small ($\rm \aplt 2$\,pc) star clusters at ages $\apgt 20$\,Myr,
although we argue that proper inclusion of tidal effects would
mitigate this discrepancy.

The stars in the simulations are initially distributed as single
stars, binaries and triples. During the evolution, multiple systems
containing more than 4 stars form as a result of strong dynamical
interactions among primordial single stars, binaries and triples.
After the first few million years the fraction of higher-order systems
remains roughly constant.

Our simulation models form significant numbers of hierarchical systems
consisting of four or more stars when compared to similar simulations
without primordial triples. We measure the fractions of systems
containing up to 6 stars, and find a steady increase in their number.
After the start of the simulation the number of higher-order multiples
rapidly increases up to an age of about 2\,Myr, after which their net
formation rate drops by about a factor of 20.  The number of multiples
consisting of 4 stars increases at about twice the rate as those
containing 5 stars, which again increase at about twice the rate of
system with 6 stars.

Among the higher-order multiples that form in our simulations, the
strictly hierarchical systems are the least likely to survive.  This
is most easily seen in the quadruples, among which the configuration
consisting of two binaries orbiting one another are by far the most
common.  Among the quintuples the most likely stable configuration is
a triple star orbited by a binary.  Most sextuple systems consist of
two triples orbiting one another.

The relative frequency of stable hierarchies in our simulations is
generally comparable to those observed in MSC, but with some notable
exceptions.  In four observed systems, a massive binary is
hierarchically orbited by three single stars.  In our simulations such
sextuple systems are extremely rare (relative frequency $\ll 1$\%).
On the other hand, a sextuple system consisting of three binaries
orbiting one another as a hierarchical triple is quite common in our
simulations, but none are observed in the population of multiple
systems in the solar neighbourhood.  These interesting differences may
originate from observational selection effects or from specific
choices in our initial conditions.

We also present the principal interactions in which multiple stars are
created and destroyed.  Reactions in which single stars ($\sim 50$\%)
or triples ($\sim 30$\%) participate are most common, simply because
such systems are most abundant.  Most sextuple systems are formed from
an interaction between two triple systems, and most quadruples form
from a triple and a single star.  The actual relative importance of
various reactions in the network may be quite different for a
different choice of initial conditions.  The initial fractions of
stars, binaries and triples may play a crucial role here, and a
shallower initial density profile may boost the number of high order
multiples and may even allow for systems containing larger numbers of
stars.  We expect, however, that multiplicities of 4 or higher will
remain relatively rare compared with systems consisting of fewer
stars.

\section{Acknowledgments}
It is a pleasure to thank John Fregeau, Douglas Heggie, Piet Hut, Jun
Makino and Andrei Tokovinin for discussions.  This research is
supported financially by NWO (via grants \#635.000.001 and
\#643.200.503), the Kapteyn fund, the Netherlands Research School for
Astronomy (NOVA), the Royal Academy of Arts and Sciences (KNAW), and
the NASA Astrophysics Theory Program (NNG04GL50G).  Parts of the
manuscript were completed during a visit (by SPZ and SLWM) to the
Kavli Institute for Theoretical Physics at UC Santa Barbara, supported
in part by the National Science Foundation under Grant
No. PHY99-07949.
%
%\nocite{*}
%\bibliographystyle{mn2e}
%\bibliographystyle{/home/spz/Latex/lib/inputs/aabib}
%\bibliography{references}

%
\bsp
\label{lastpage}
\end{document}